\documentclass[mathleft
]{an}
\usepackage{graphicx}
\usepackage{times}
\begin{document}

\Pagespan{001}{}
\Yearpublication{2017}%
\Yearsubmission{2016}%
\Month{11}%
\Volume{000}%
\Issue{00}%

\title{Tidal disruption of stars by supermassive black holes -- XMM-Newton highlights and the next decade}

\author{S. Komossa\inst{1,2}\fnmsep\thanks{Corresponding author: on leave of absence from MPIfR, Bonn, Germany;
  \email{astrokomossa@gmx.de}\newline}
}
\titlerunning{TDEs with XMM-Newton}
\authorrunning{S. Komossa}
\institute{National Astronomical Observatories, Chinese Academy of Sciences, Beijing, China
\and 
QianNan Normal University for Nationalities, Longshan Street, Duyun City of Guizhou Province, China}

\received{11 Nov 2016}
\accepted{11 Nov 2005}
\publonline{later}


\abstract{%
This article provides a summary of XMM-Newton highlights on stellar tidal disruption
events. First found with ROSAT,
ongoing and upcoming sky surveys will detect these events in the 1000s. In X-
rays, tidal disruption events (TDEs) provide us with powerful new probes of
accretion physics under extreme conditions and on short timescales and of relativistic
effects near the SMBH, of the formation and evolution of disk winds near
or above the Eddington limit, 
and of the processes of high-energy emission from newly launched radio jets. 
TDEs serve as signposts of the presence of dormant, single black holes
at the cores of galaxies, and of binary black holes as well, since TDE lightcurves 
are characteristically different in the latter case. 
XMM-Newton has started to contribute to all of these topics, and a 
rich discovery space is opening up in the next decade. }

\maketitle

\section{Introduction}

Supermassive black holes (SMBHs) can violently disrupt stars in their vicinity. 
The subsequent accretion of stellar material produces spectacular flares 
in the X-ray sky, and therefore can reveal the actual presence of the SMBH,
even and especially if the black hole itself is usually dormant and does not harbour a
long-lived accretion disk. Flares from tidally disrupted stars therefore represent
independent signposts of the presence of SMBHs in non-active
galaxies (e.g., Rees 1988) which are otherwise hard to detect,
 and provide us with a new route of estimating masses and spins of SMBHs.    
Because single stars are disrupted in the {\em immediate} vicinity of a SMBH,
this method of black hole detection is very complementary  to another well-established
detection method, which makes use of the motions of large ensembles of
stars and gas at much larger distances from the SMBH.  

Theoretical foundations of the stellar disruption process were laid in the 1970s, followed
by early suggestions, how and where TDEs could play a role in astrophysical
environments, including speculations about their contribution to fuelling active
galactic nuclei (AGN), growing black holes, and powering gamma-ray bursts.     

A star approaching a SMBH will feel increasingly strong tidal forces. Once the 
self-gravity of the star can no longer withstand those forces, the star is ripped
apart (Hills 1975). This happens at the tidal radius, 

\begin{equation}
r_{\rm t} \simeq 7 \times 10^{12}\,{\bigg{(}}{M_{\rm BH}\over {10^{6} M_\odot}}{\bigg{)}}^{1 \over 3}
    {\bigg{(}}{M_{\rm *}\over M_\odot}{\bigg{)}}^{-{1 \over 3}} {r_* \over r_\odot}~{\rm cm}\,. 
\end{equation}

Only a fraction of the stellar material will eventually be accreted,
while the remainder is unbound and escapes the system.
The bound stellar material follows Keplerian orbits of high eccentricity 
and returns to pericenter at a rate

\begin{equation}
dM/dt \propto t^{-5/3}. 
\end{equation}

Major observational signatures of TDEs, which aid their identification, include 
(1) a flare of electromagnetic radiation at high peak luminosity, (2) from the core
of an otherwise quiescent, inactive galaxy, (3) with a lightcurve which characteristically
shows a rapid rise followed by a decline on the timescale of months to years. (4) If the
stellar material is accreted as quickly as it returns to the SMBH after it was
initially spread over a range of Keplerian orbits, the X-ray lightcurve is expected
to decline as $t^{-5/3}$.

First events were identified in the 1990s (e.g., Komossa \& Bade 1999)
in the course of the soft X-ray 
all-sky survey carried out with
the ROSAT satellite, in form of luminous, very soft,
giant-amplitude X-ray flares from otherwise quiescent galaxies, matching well all
order-of-magnitude predictions of tidal disruption theory (Rees 1988), including
an overall lightcurve decline consistent with a t$^{-5/3}$ law.     

More recently, events have been found at other wavebands 
(review by Komossa 2015); in the UV, optical, 
and at hard X-rays, and with follow-up detections in the NIR and radio 
bands. Some of these come with transient optical emission lines from hydrogen,
helium and iron coronal transitions. 
About 40 events have been identified by now.  

From a theoretical point of view, the simulation of the tidal disruption and accretion of a star,
including predictions of the detailed observational charateristics at any given
time of the evolution (multi-waveband spectra and lightcurves) poses many challenges
(review by Lodato et al. 2015). 
Substancial progress in recent years has 
addressed the stellar evolution, including the early stages of stellar deformation, 
the actual disruption, accretion and ejection of material, the development of a disk wind, 
and the evolution and late stages of the accretion phase, and for different types of stars
in different orbits, and for single stars as well as binary stars, and for spinning
and non-spinning black holes
(e.g., Bonnerot et al. 2016, Chen et al. 2016, Coughlin et al. 2016a, Coughlin et al. 2016b,  
Franchini et al. 2016, Hayasaki et al. 2016, Liu et al. 2016, Lu et al. 2016, MacLeod et al. 2016, 
Mainetti et al. 2016a, Metzger \& Stone 2016,
Ricarte et al. 2016, Roth et al. 2016, Sadowski et al. 2016, and references therein;
review by Lodato et al. 2015).

X-ray observations of TDEs probe the inner parts of the accretion disk,
and are of outstanding importance for our understanding of accretion physics
under extreme conditions, and on very short timescales, including the evolution
of the accretion disk through several accretion regimes from super- to very sub-Eddington
on the timescale {\em of only years}.

\section{Highlights from XMM-Newton}

The X-ray mission XMM-Newton was launched in 1999 (Jansen et al. 2001). 
The satellite is equipped with 
CCD arrays, sensitive between (0.2-12) keV, with moderate spectral resolution,
and a spatial resolution of 6'' (Turner et al. 2001, Str{\"u}der et al. 2001). 
High-resolution spectral information is provided
by the Reflection Grating Spectrometer, RGS (den Herder et al. 2001). 
Additional optical/UV information is delivered by an onboard small 
optical/UV telescope, the Optical Monitor (OM; Mason et al. 2001). 

In addition to pointed observations, XMM-Newton carries out a slew survey, which 
registers X-ray photons along the path while moving from one target to another one,
performing a mini-survey of the sky that way. 
Further, a mode of Target of Opportunity (ToO) observations allows for rapid follow-ups 
of flaring and transient sources, with a repsonse time on the order of days. 

This section provides an overview of TDE results obtained with XMM-Newton, including
the discovery of new events, and follow-ups of remarkable events first identified
at other wavebands. These observations have added substancially to our understanding
of the early evolution of TDEs, and the underlying accretion and jet physics.

\subsection{First hard X-ray follow-up of a TDE: spectral hardening of RXJ1242--1119}

The TDE RXJ1242--1119 was first identified by its luminous X-ray flare ($L_{\rm x, peak} > 10^{44}$ 
erg/s) detected with the 
ROSAT X-ray observatory (Komossa \& Greiner 1999). It is located in an inactive host galaxy at redshift
$z$=0.05. It was the first TDE to be followed-up at hard X-rays with Chandra and XMM-Newton, and the
first, for which an XMM-Newton (CCD) spectrum could be obtained (Komossa et al. 2004).
While its initial ROSAT X-ray spectrum was exceptionally soft ($\Gamma_{\rm X} \simeq -5$, or 
$kT_{\rm bb} = 0.0 6$ keV), 
it had hardened signifcantly ($\Gamma_{\rm X} = -2.5$) when observed
with XMM-Newton (Komossa et al. 2004, Halpern et al. 2004), 
perhaps implying the formation of a disk corona, which was not present in the very early
stages of TDE evolution, and/or a change in accretion mode. A similar hardening of the 
X-ray spectrum was observed on shorter timescales
for NGC 5905 (Komossa \& Bade 1999) and more recently in other TDE candidates 
(e.g., Nikolajuk \& Walter 2013).   
These follow-up observations also revealed a very high amplitude of X-ray decline of RXJ1242-1119, exceeding
a factor of 1500 on the timescale of a decade, implying a strong change in the accretion conditions 
on relatively short timescales. Such a high amplitude of variability has not  been observed in any AGN. 

At high-state, the accretion rate of RXJ1242--1119 was below the Eddington limit, given its black hole
mass of $M_{\rm BH} \simeq 10^8$ M$_{\odot}$, estimated from the XMM-Newton Optical Monitor (OM)
blue magnitude, and applying SMBH -- host scaling relations (Komossa et al. 2004). 
Based on the observed lightcurve, the integral
$\int L(t) dt$ provides us with a lower limit on the total emitted energy of the 
event, $E > 1.6 \times 10^{51}$ ergs. This implies a lower limit on the amount of stellar material
which was accreted, $M_* > 0.01 \eta_{\rm 0.1}^{-1}$ M$_{\rm \odot}$, where $\eta = 0.1 \eta_{\rm 0.1}$
is the efficiency.

\subsection{Discovery of new soft X-ray TDEs based on the XMM-Newton slew survey: early-phase lightcurves}

A dedicated search for new TDEs near maximum, including (rapid) multi-wavelength follow-ups, is being
carried out based on the XMM-Newton slew survey (Esquej et al. 2007, 2008, 
Saxton et al. 2012, 2015, 2017).  
Events near high-state are identified by cross-correlation with the ROSAT database, and TDE candidates
with soft spectra and high amplitudes of variability are selected for follow-ups, including optical
spectroscopy in order to confirm non-active host galaxies. 

Among these events, SDSSJ120136.02+300305.5 has the best-covered first-year X-ray lightcurve.
It was detected in an XMM-Newton slew in 2010 with a flux 56 times higher than an
upper limit from ROSAT, with a high peak luminosity of $L_{\rm x} = 3 \times 10^{44}$ erg/s
(Saxton et al. 2012). 
Its X-ray lightcurve shows an overall decline by a factor $\sim300$ over 300 days, superposed
onto which is strong variability, reaching up to a factor of $>$50, when the source is 
no longer detected by Swift. The host galaxy, at redsift $z$=0.146, is inactive, and no
broad Balmer lines or coronal lines have been detected in response to the X-ray flare,
in spectra taken 12 days and 11 months after discovery. 
Unlike other TDEs, SDSSJ120136.02+300305.5 does not show an X-ray spectral hardening with time, but
remains relatively soft, and its XMM-Newton X-ray spectrum is not well fit by black-body-like thermal
emission, neither by a single powerlaw.  Instead, the spectral shape is modelled well by
either a broken powerlaw, or Bremsstrahlung emission of $kT$ = 390 eV.  
No radio emission was detected from
SDSSJ120136.02+300305.5 by the VLA between 1.4 and 8.3 GHz, 
implying that no powerful radio jet was launched during this event. 

The candidate TDE from NGC\,3599, again identified during an XMM-Newton slew
(Esquej et al. 2007, Esquej et al. 2008),
is interesting because of its relatively long-lived X-ray emission, implying either
a long-lived TDE, or else a permanent low-luminosity AGN in this LINER galaxy (Saxton et al. 2015).  
X-ray emission from NGC\,5905 was bright and at similar levels{\footnote{note that the
observed peak luminosity from this event was relatively low, $L_{\rm x} \approx 10^{41}$ erg/s,
making this the lowest-luminosity event among the TDEs and candidates}} 
during two XMM-Newton slews from 2002 and 2003,
18 months apart. Its soft X-ray
flux was a factor of $\sim$150 higher than an upper limit from ROSAT,
and declined after the second slew observation in follow-ups with XMM-Newton, Swift,
and Chandra by a factor of 100. 
While classical TDEs are expected to show a fast rise to maximum
and a decline on the timescale of months, driven by quick circularization (due to stream-stream
collisions and shocks) 
and accretion of the stellar debris (e.g., Rees 1988), recent  
simulations have predicted a class of longer-lived TDEs (e.g., Guillochon \& Ramirez-Ruiz 2015),
which arise when circularization of the stellar material is delayed. 
As discussed by Saxton et al. (2015), the flare from NGC\,3599 may represent 
such a delayed TDE. Alternatively, it may be
explained by a thermal instability in the accretion disk,
which propagates through the inner region at the sound speed, causing an increase of
the local accretion rate. The latter scenario predicts episodes of repeat flaring
on the timescale of decades, as the inner disk is emptied and refilled, perhaps
similar to what has been seen in the Seyfert galaxy IC\,3599 (Grupe et al. 2015) which exhibited two
giant X-ray outbursts $\sim$20 yrs apart.

Identifying new TDEs early in XMM-Newton slews provides us with deep insight on the very
early phases of TDE evolution, and is important for quickly 
triggering multi-wavelength follow-up observations.

\subsection{Search for new TDEs based on the XMM-Newton archive} 

Several TDEs have been identified with XMM-Newton, based on either dedicated archival
searches or the analysis of clusters of galaxies or chance detections (Cappelluti et al. 2009, 
Maksym et al. 2010, 2013, 2014, Khabibullin \& Sazonov 2014, Donato et al. 2014, 
Lin et al. 2011, 2015, Mainetti et al. 2016b). 
In addition to studying the properties of single events in different host 
galaxies and environments, such searches are also important for estimating TDE rates.  

TDXFJ134730.3-325451 was serendipitously identified as an X-ray transient source in the 
cluster of galaxies Abell 3571 (Cappelluti et al. 2009). 
It was brightest during ROSAT observations, but was
only recognized as transient source in later XMM-Newton and Chandra observations of 
the cluster Abell 3751, when it was fainter by a factor 650 than at its highest state. 
While a high-resolution optical spectrum of the host galaxy LEDA\,095953 is not yet
available, photometry indicated an inactive host, and the high amplitude of variability
and the high peak luminosity of $\sim 10^{43}$ erg/s (0.3-2.4 keV)
made this a good TDE candidate. 
Based on a second TDE identified in a cluster of galaxies, Maksym et al. (2011) estimated
a TDE rate of $1.2 \times 10^{-4}$ events per galaxy per year. 

In the course of a dedicated search for TDEs in clusters of galaxies, Maksym et al. (2013)
and Donato et al. (2014) reported the detection of an X-ray TDE candidate near the core of 
Abell 1795. The likely host galaxy is a low mass galaxy at 
redshift $z=0.062$ with $M_{\rm BH} < 10^6$ M$_{\odot}$. 

3XMMJ152130.7+074916 was 
serendipitously detected in an XMM-Newton observation in 2000
as an ultra-soft X-ray transient source.
Its location is consistent with the center of the galaxy SDSS J152130.72+074916.5 at 
$z$ = 0.179, located in a cluster of galaxies.
   The X-ray spectrum is consistent with thermal disk 
emission of temperature 0.17 keV and 
a rest-frame (0.24-11.8 keV) luminosity of $5 \times 10^{43}$ erg/s, 
subject to a high-velocity warm absorber (Lin et al. 2015).

\subsection{A candidate sub-milli-parsec binary SMBH identified from a TDE lightcurve}

If a TDE occurs in binary supermassive black hole system, the binary will imprint its
presence on the TDE lightcurve. The secondary black hole will perturb the 
stream of stellar material, causing 
temporary interruptions of the accretion flow on the primary,
leading to epochs of characteristic deep dips and recoveries 
in the decline lightcurve (Liu et al. 2009, Ricarte et al. 2016). This signature
has been observed in the lightcurve of the TDE from SDSSJ120136.02+300305.5.
While its X-ray lightcurve is consistent with an overall decline, 
about a month after high-state, the X-rays fade by a factor $>$50
within a week, and remain undetected between day 27 and day 48 
after discovery. Then, the X-rays are bright again,
followed by a second disappearance. 
No excess absorption was detected in X-rays, and no jet was seen 
in radio follow-ups. 
Detailed simulations have shown that the lightcurve 
of SDSSJ120136.02+300305.5 can be reproduced well 
with a model of a SMBH binary of mass ratio $q \simeq$ 0.1 at 
0.6 milli pc separation (Liu et al. 2014). 

Tightly-bound SMBH binaries like this one, which have already overcome 
the ``final parsec problem'', are prime sources of gravitational wave 
radiation once the two black holes coalesce. 
For a binary SMBH system in SDSSJ120136.02+300305.5 with a primary mass 
of $M_{\rm BH} = 10^7$ M$_{\odot}$ and mass ratio $q$=0.08, an orbital timescale of 150 d and 
with orbital eccentricity $e$ = 0.3 (Liu et al. 2014), the lifetime due to emission
of gravitational waves is $t_{\rm GW} \simeq 2 \times 10^6$ yr.   

Future observations of TDE lightcurves provide us with a powerful new tool
of searching for compact binary SMBHs, well below spatial separations
of parsecs, and in {\em inactive} galaxies which do not harbour long-lived AGN,
and where binary SMBHs are especially difficult to detect by other means
(e.g., Komossa \& Zensus 2016).

\subsection{Follow-ups of the first jetted TDE discovered by Swift}

The event Swift\,J1644+57, discovered with Swift in 2011, differs from previous TDEs,
in that it was detected at hard X-rays and was 
accompanied by strong (beamed) radio jet emission 
(e.g., Burrows et al. 2011, Bloom et al. 2011, Zauderer et al. 2011, 
Berger et al. 2012, Zauderer et al. 2013,
Levan et al. 2011, Mangano et al. 2016, Yang et al. 2016). 
Swift\,J1644+57 was first detected as a bright X-ray source by the Swift 
Burst Alert Telescope (BAT) in
March 2011. Its (isotropic) peak luminosity exceeded 10$^{48}$ erg/s. 
Its X-ray lightcurve shows a
general downward trend, on which very rapid, very high-amplitude variability is superposed, 
changing on timescales as fast
as 100s. The host galaxy at redshift $z = 0.35$ does not show signs of permanent optical
AGN activity. Low-ionization emission lines indicate starformation activity. 
The event is accompanied by unresolved and variable radio emission.  
The observations have been interpreted as the rapid onset of a powerful 
jet following the tidal disruption of a star.
The event has the best-covered X-ray lightcurve of any TDE todate,
and has motivated a large number of follow-ups and theoretical studies (review by
Komossa 2015), with an emphasis on the question of jet launching under TDE
conditions, and the role of magnetic fields (e.g., Tchekhovskoy et al. 2014).
 
XMM-Newton has observed Swift\,J1644+57 multiple times during the first year
after discovery (Reis et al. 2012, Castro-Tirado et al. 2013, 
Gonzales-Rodriguez et al. 2014). 
The first of these follow-ups was performed a few days after the BAT trigger.
Evidence for
a possible QPO at 5 milli Hz was reported by Reis et al. (2012).
The QPO is only present in hard X-rays (2-10 keV)
and only in one of the XMM-Newton observations (the one taken $\sim$19 d
after the BAT trigger), as well as in Suzaku data. 
Reis et al. suggested an origin in the inner accretion disk 
(see also Abramowicz \& Liu 2012),
while an association with turbulence in the accretion flow or resonances in
the relativistic jet could not be ruled out. 

Recently, Kara et al. (2016) reported the detection of a reverberation signal in
X-rays, arising when gravitationally redshifted photons from the iron K$\alpha$
emission line  reflect off the inner accretion disk. 
These observations indicate, that the bulk of the X-ray emission at early times
is dominated by the accretion disk. 
From the reverberation time lag, a SMBH mass of a few 10$^6$ M$_{\odot}$
was estimated, confirming a highly super-Eddington accretion flow.

Swift continued to monitor Swift\,J1644+57 at daily to weekly 
time intervals, to follow the lightcurve decline. 
After $\sim$1.5 yr, the X-rays from Swift\,J1644+57 
suddenly dropped by a factor $\sim$100, and have remained
faint ever since (at $L_{\rm x} = 5 \times 10^{42}$ erg/s,
with $\Gamma_{\rm x} \sim 1.9$),  
 no longer detected by Swift, but only in 
deeper observations of Chandra and XMM-Newton (Zauderer et al. 2013, Mangano
et al. 2016, Levan et al. 2016). The reason for this 
dramatic change is not yet fully understood.
The low-state X-rays may arise from the forward shock associated with the radio
jet (Zauderer et al. 2013), but have not yet faded further so far, while the 
radio emission does (Yang et al. 2016). 
Cheng et al. (2016) explored the possibility that the low-state X-ray emission
arises from Thomson scattering of the primary X-rays off surrounding plasma.   
A persistent low-luminosity AGN is
an unlikely explanation for the long-lived low-state X-rays, since the X-ray luminosity
is still significantly higher than expected from the faint, low-mass host galaxy
(Levan et al. 2016). The low-state X-ray spectrum is rather hard, inconsistent
with thermal emission from the accretion disk (Zauderer et al. 2013), 
but possibly due to Comptonized emission from a disk corona (Mangano et al. 2016).   

This and future observations of jetted TDEs provide us with a completely new probe
of the early phases of jet formation and evolution,
and the jet-disk connection, in an otherwise quiescent environment
without past X-ray and radio-AGN activity.

\subsection{Detection of absorption lines in a TDE X-ray spectrum}

The first X-ray grating spectrum of a TDE was presented
recently (Miller et al. 2015). 
ASASSN14li at redshift $z=0.02$ was first identified based on its optical variability
in a supernova search survey (Jose et al. 2014, Holoien et al. 2016). It is accompanied
by luminous X-ray, and faint, variable radio emission (van Velzen et al. 2016,
Alexander et al. 2016). 
  
The high-resolution RGS X-ray spectrum of ASASSN14li revealed 
the presence of several narrow absorption lines, superposed on soft, thermal
continuum emission ($kT = 0.05$ keV). Velocity shifts of few 100 km/s 
imply matter in outflow. The material is highly ionized with 
an ionization parameter on the order of log $\zeta \simeq 4$, 
and a column density of the ionized gas of order few 10$^{21}$ cm$^{-2}$. 
 
Narrow-line variability (mostly in the outflow velocity) was
used to argue for a location of the outflowing material close to
the SMBH. Miller et al. (2015) suggest a rotating wind from the inner, super-Eddington, 
nascent accretion disk, or a filament of the disrupted star itself, 
as most likely origin of the absorbing gas. 

This observation has opened up a new window into studying accretion-driven winds
powered by TDEs in X-rays. Future absorption line spectroscopy will tighly constrain
the physical properties, abundances, and velocity fields of newly launched disk winds,
and the stellar material itself.

\section{Future opportunities and the next decade with XMM-Newton}

Chances of discovering new TDEs with XMM-Newton arise from the continuing
dedicated search for transients in the XMM-Newton slew survey, in the XMM-Newton archive, in clusters
of galaxies, and from serendipitous detections. Chances for 
deep spectroscopic follow-ups of TDEs with XMM-Newton will additionally 
open up, when new TDEs are identified 
in ongoing or upcoming sky surveys. These include current surveys such as those carried out with 
PTF, Pan-STARRS, and ASAS-SN in the optical, or Swift and MAXI in X-rays and $\gamma$-rays, as well
as future surveys like with LSST and with SKA. In X-rays, several
proposed missions like the Chinese {\sl Einstein Probe} will carry out dedicated transient 
searches, and are likely to have an overlap in time with XMM-Newton.  

In the future, the performance of XMM-Newton might be further enhanced by aiming
at an even more rapid response after the discovery of new TDEs, by increasing
the frequency and depth of (rapid) follow-up X-ray observations if interesting spectral features
in emission or absorption are present, by setting up dedicated {\sl{large programs}}
and/or automated follow-ups of
transients or TDEs in particular, and by increasing the mutual agreements with other observatories 
for joint multi-wavelength follow-ups, for instance in the radio regime with the VLA, EVN or FAST.           

In summary, deep X-ray observations with XMM-Newton provide us with a unique 
chance of probing accretion physics down to the last stable orbit and under
extreme conditions (near-Eddington up to hyper-Eddington accretion rates).
They allow us to follow the evolution
of disk winds and coronae, search for relativistic 
(precession) effects in the Kerr metric, perform iron-line reverberation mapping, 
estimate BH spin, carry out absorption/emission-line 
spectroscopy of ionized matter in outflow (either stellar debris or accretion disk winds),
and study the jet-disk coupling and jet evolution in jetted events.                                                                              
All of these are now within reach and first examples of these processes have been found with
XMM-Newton recently, opening up a rich discovery space
in the next decade.




\end{document}